\documentclass[referee]{jfm}

\usepackage{upmath}
\usepackage{amstext}
\usepackage{amsmath}%
\usepackage{amsfonts}%
\usepackage{amssymb}%
\usepackage{amsbsy}
\usepackage{amsgen}
\usepackage{natbib}
\usepackage{graphicx}
\usepackage{color}

\title[The motion of a sphere around an obstacle]{Trajectory and distribution of suspended 
non-Brownian particles moving past a fixed spherical or cylindrical obstacle}
\author[Sumedh R. Risbud and German Drazer]{By\ns S\ls U\ls M\ls E\ls D\ls H\ns R.\ns R\ls I\ls S\ls B\ls U\ls D\ns 
\and G\ls E\ls R\ls M\ls A\ls N\ns D\ls R\ls A\ls Z\ls E\ls R}
\affiliation{Department of Chemical and Biomolecular Engineering\\[\affilskip]Johns Hopkins University}
\date{\today}
\begin{document}
\maketitle

\begin{abstract}
We investigate the motion of a suspended non-Brownian sphere past a fixed cylindrical or spherical obstacle in the limit of zero Reynolds number for 
arbitrary particle-obstacle aspect ratios. We consider both a suspended sphere moving in a quiescent fluid under the action of a uniform force as well 
as a uniform ambient velocity field driving a freely suspended particle. We determine the distribution of particles around a single obstacle and solve 
for the individual particle trajectories to comment on the transport of dilute suspensions past an array of fixed obstacles. First, we obtain an 
expression for the probability density function governing the distribution of a dilute suspension of particles around an isolated obstacle, and we show 
that it is isotropic. We then present an analytical expression -- derived using both Eulerian and Lagrangian approaches -- for the minimum 
particle-obstacle separation attained during the motion, as a function of the incoming impact parameter, i.e. the initial offset between the line of 
motion far from the obstacle and the coordinate axis parallel to the driving field. Further, we derive the asymptotic behaviour for small initial 
offsets and show that the minimum separation decays exponentially. Finally we use this analytical expression to define an effective hydrodynamic surface 
roughness based on the net lateral displacement experienced by a suspended sphere moving past an obstacle.
\end{abstract}

\maketitle

\section{Introduction}
The interaction between suspended particles and obstacles encountered in their flow is essential for the 
understanding of the transport of particulate suspensions in natural porous systems as well as in engineered 
porous media used in different applications. In porous media, the behaviour of individual particles at the 
`pore-scale' determines the migration, dispersion and ultimate fate of suspended particles \citep{lee1999}. 
Lagrangian methods in filtration theory, for example, rely on calculating the trajectory of individual particles 
in simplified representations of the pore space \citep{spielman1977,adamczyk1989,ryan1996,jegatheesan2005}. 
Particle capture by single cylindrical and spherical collectors is investigated to understand filtration in 
fibrous media and packed beds, respectively \citep{spielman1977}. In many filtration methods, such as deep bed 
filtration, the particles are typically smaller than the characteristic scale of the pores, and get collected or 
deposited onto the obstacles in their path. In order to gain insight into these systems, several studies have 
focused on the trajectories followed by small particles near large spherical or cylindrical collectors 
\citep{adamczyk1981, adamczyk1989, adamczyk1989a, goren1971, gu2002, li2002}. Trajectory studies have also shown 
that not only particle capture but transit times may also be dominated by single particle-obstacle interactions 
\citep{lee1999}. Other studies modelling particle transport in porous media have focused on the similar problem 
of the trajectory followed by individual particles as they move through narrow channels, such as in the 
constricted-tube model \citep{burganos1992, burganos2001,chang2003}. Studies of particle migration, dispersion and 
capture using Eulerian methods also rely on calculating the detailed particle-fibre or particle-grain interactions 
in model porous media \citep{koch1989,phillips1989,phillips1990,shapiro1991,nitsche1996}. In this work, we investigate 
the properties of particle trajectories as the particles move past a fixed spherical or cylindrical obstacle driven by 
either a constant force or a uniform velocity field in the presence of particle-obstacle hydrodynamic interactions and 
for arbitrary particle-to-obstacle size ratios.

Recently, with the advent of microfluidic technology, there is a renewed interest in the understanding and 
characterization of the motion of individual particles past solid obstacles. The possibility of designing the 
stationary media with nearly arbitrary geometry and chemistry, with features of micron and sub-micron size, has led to 
microfluidic separation techniques that are primarily based on the species-specific particle-obstacle interactions. In 
deterministic lateral displacement, for example, a mixture of suspended particles driven through a two-dimensional (2D) 
periodic array of cylindrical obstacles spontaneously fractionates as different species migrate in different directions 
\citep{huang2004}. The migration angle depends on the particle-obstacle interactions, and can be accurately described 
based on the trajectory followed by individual particles as they move past a fixed obstacle 
\citep{frechette2009,balvin2009,herrmann2009,bowman2012}. Particle-obstacle interactions are also essential in separation 
methods based on ratchet effects induced by asymmetric arrays of obstacles \citep{li2007}. Other methods, such as pinched 
flow fractionation \citep{yamada2004} and some particle 
focusing systems \citep{hewitt2010,xuan2010} can also be investigated from the perspective of particle-obstacle 
interactions and the effect that both hydrodynamic and non-hydrodynamic forces have on the trajectories followed by 
different species \citep{luo2011}. In all cases, the individual particle trajectories or the asymptotic distribution 
of suspended particles as they are driven past spherical or cylindrical obstacles contains the information 
necessary for calculation of the relevant average transport properties, such as migration speed and angle. However, 
an analytical treatment describing these two aspects of particle motion has not been developed. In this 
work, we provide such an analysis, within the purview of low Reynolds number hydrodynamics, wherein Stokes' 
equations are assumed to describe the motion of the fluid, and particle inertia is neglected. Further, we consider 
the non-Brownian regime, since in many applications -- separation devices in particular 
-- high throughput is advantageous, leading to a routine occurrence of high P\'eclet numbers and the motion of the 
particles can be approximated as deterministic.

In addition, 
the results derived here are of relevance to suspension rheology and flows, specially in active micro-rheology, where a 
probe particle is driven through a quiescent suspension of particles. In turn, we shall show that it is possible to 
extend analytical results established in suspension rheology to the motion of a spherical particle driven past a fixed 
obstacle. 
In suspension flows, each individual particle is moving through a random distribution 
of other spheres. Furthermore, in the dilute approximation,  the analysis of the relative motion (and distribution) 
between any two spheres of the suspension suffices to characterize its properties. Such a relative motion is 
analogous to the problem of a sphere moving with respect to a fixed obstacle in -- say, for example -- a microfluidic 
device. (Note that in the latter case the dilute approximation assumes not only a dilute suspension but also that the 
distance between obstacles is sufficiently large.) Therefore, before 
proceeding, it is convenient to briefly review some of the most relevant results concerning two-particle interactions 
in the limit of zero Reynolds and infinite P\'{e}clet numbers. \citet{batchelor1972a} first obtained the pair distribution 
function of spheres in a sheared suspension. Batchelor later extended his work by deriving the pair-distribution function 
in the case of a sedimenting polydispersed suspension of spheres \citep{batchelor1982, batchelor1982a, batchelor1983}.
\citet{davis1992} used the resulting expression for the pair distribution function to analyze the hydrodynamic diffusion 
of a sphere sedimenting through a dilute suspension of neutrally buoyant particles. \citet{almog1997} further employed 
the same pair distribution function to study the apparent viscosity experienced by a sphere moving through a quiescent 
suspension. They considered both the micro-rheological setting of a `falling ball viscometer' as well as a non-rotating 
sphere moving with a prescribed velocity through the suspension. \citet{khair2006a} studied the motion of a single Brownian 
particle through a suspension, and extended the analysis of \citet{batchelor1982} to obtain the pair-probability distribution 
function for finite P\'{e}clet numbers. 

In this work, we consider the trajectory followed by a suspended particle driven by a constant 
external force (e.g. buoyancy force) or a uniform velocity field as it moves past a fixed sphere or cylinder in the limit 
of zero Reynolds and infinite P\'eclet numbers. We derive an expression for the minimum particle-obstacle 
separation attained during the motion, as a function of the incoming impact parameter between the sphere trajectory 
and the centre of the obstacle (offset $b_{in}$ in figure \ref{fig:schematicRepresentation}). The minimum distance 
between the particle and obstacle surfaces, that would result from a trajectory determined solely based on hydrodynamic 
interactions, dictates the relevance of short-range non-hydrodynamic interactions such as van der Waals forces, surface 
roughness (solid-solid contact), etc. In fact, the resulting scaling relation derived for small $b_{in}$ shows that 
extremely small surface-to-surface separations would be common during the motion of a particle past a distribution of 
periodic or random obstacles. This highlights the impact that short-range non-hydrodynamic interactions could have in the 
effective motion of suspended particles. In this context, we further demonstrate that the particle attains smaller 
surface-to-surface separations from the obstacle when confined in a channel, with walls parallel to the 
direction of motion of the particle. We also calculate the distribution of particles around a fixed obstacle 
in the dilute limit, by extending the existing analytical treatment in the case of sheared suspensions. We show that the 
steady state distribution is isotropic, which is a somewhat surprising result given the anisotropy induced by the driving field, 
but it is analogous to the distribution obtained in the cases of sheared and sedimenting suspensions investigated by 
Batchelor \citep{batchelor1972a,batchelor1982}. The steady state distribution of particles provides the necessary 
information to obtain macroscopic transport properties, such as the average velocity of the suspended particles, from 
the pore-scale point-wise velocity \citep{brenner1993}.

The organisation of the article is as follows: in the next section we formulate the problem, describe the relevant 
geometries and briefly summarise the available results for the mobility functions corresponding to these geometries. 
In \S\ref{sec:steadyStateDistrib} we derive the particle-obstacle pair-distribution function in terms of the mobility 
functions, and consider the limiting cases of nearly touching particle-obstacle pairs (\S\ref{subsec:smallXiPDF}) 
as well as widely separated pairs (\S\ref{subsec:largeXiPDF}). In \S\ref{sec:trajectoryAnalysis} we derive an 
expression for the minimum surface-to-surface separation attained during the course of motion of a particle around an 
obstacle. The derivation follows two distinct paths outlined in \S\ref{subsec:trajAnalEuler} and 
\S\ref{subsec:trajAnalLagrange}; the former follows an Eulerian approach by deriving the expression using the 
probability distribution obtained in the preceding section, while the latter arrives at the same expression using a 
Lagrangian approach by  calculating the particle trajectory. Then in \S\ref{subsec:smallXi}, we discuss the scaling 
of the minimum surface-to-surface separation in the limiting case of particles nearly touching the obstacle due to 
small incoming impact parameters. Finally, in \S\ref{sec:hydSurfRghnss}, we present a possible application of the 
relationship between the minimum separation and the incoming impact parameter to determine an effective hydrodynamic 
surface roughness.

\section{Model systems and formulation of the problem}\label{sec:systemAndProblem}

As discussed before, we intend to characterise the deterministic transport of a dilute suspension of 
spherical particles through an array of obstacles. Here, we refer to a dilute suspension in the sense that 
the particle-particle hydrodynamic interactions within the suspension can be neglected. 
In the case of an unbounded suspension, this approximation is accurate for particle volume 
fraction below $2\%$ \citep{batchelor1972a}. In the case of a suspension under geometric confinement, such as in a microfluidic device, 
the confinement screens the particle-particle hydrodynamic interactions, and higher volume fractions might still 
be accurately described by the dilute limit approximation. 
In addition, we assume 
that the obstacles are sufficiently separated that 
particles  interacts with only one obstacle at any 
instant. That is, we consider the situation in which a single particle 
negotiates an isolated fixed obstacle. In order to obtain the steady-state distribution of particles we shall 
assume that the incoming particles in suspension are spatially uncorrelated, i.e., they follow a uniform 
distribution far away from the obstacle. In addition, we neglect the effect of both fluid and particle inertia (zero 
Reynolds and Stokes numbers), as well as the Brownian motion of the suspended particles (infinite P\'{e}clet number). 
This leads to the consideration of pair hydrodynamic interactions between the suspended particle and a solid obstacle in 
the Stokes regime. We investigate the effect of the driving field (either a constant force or a uniform ambient velocity 
field), the obstacle type (either cylindrical or spherical), and the particle-obstacle aspect ratio. From a Lagrangian 
perspective we are interested in the trajectory followed by individual particles and, in particular, the distance of 
closest approach, $r_{min}$, as a function of the incoming impact parameter, $b_{in}$ (see figure 
\ref{fig:schematicRepresentation} for a schematic representation). From an Eulerian perspective we are interested in 
the distribution of particles around the fixed obstacles. Note that in the problem considered here, the single-particle 
probability density around a fixed obstacle is analogous to the two-particle probability or pair distribution function 
in the case of suspension flows.

\subsection{Problem geometry and symmetries}\label{subsec:geomAndSymm}

\begin{figure}
\begin{center}
\includegraphics[width=14cm]{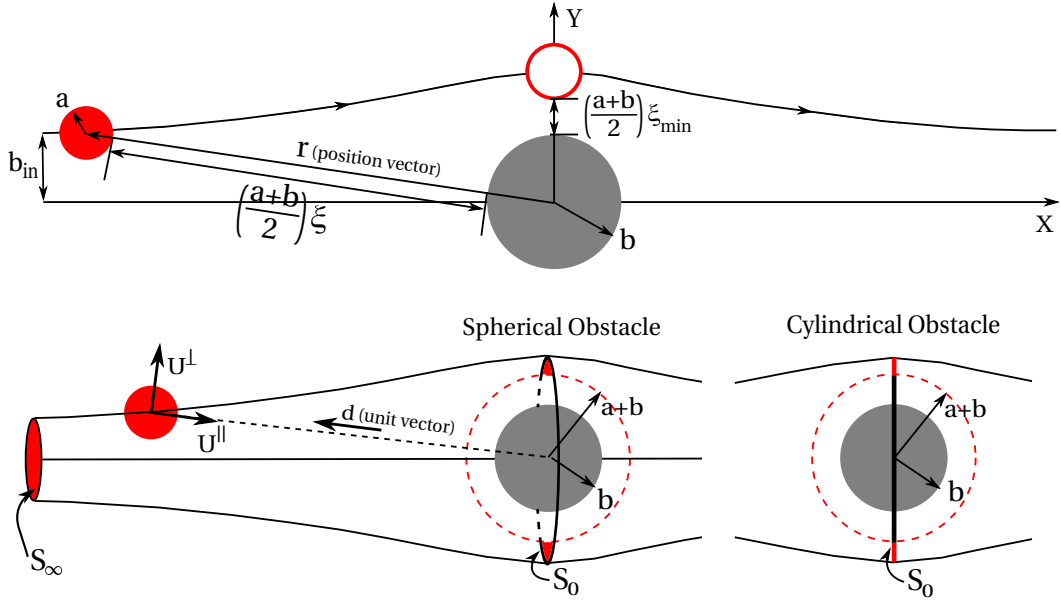}
\end{center}
\caption{Schematic representation of the problem. (top) The small circle of radius $a$
 on the left represents the moving sphere, with a corresponding incoming impact 
parameter $b_{in}$. The circle of radius $b$ with its centre at the origin of 
coordinates represents the fixed obstacle (either a sphere or a cylinder). The empty 
circle represents the position of the suspended particle as it crosses the symmetry 
plane normal to the $x$ axis. The surface-to-surface separation $\xi$ and its minimum 
value $\xi_{min}$ are also shown. 
(bottom) Representation of the conservation argument used to calculate the minimum 
separation (see \S\ref{sec:trajectoryAnalysis}). The unit vector in the radial 
direction $\mathbf{d}$ is shown, as well as the velocity components in the plane of 
motion. The surfaces over which the flux is conserved are shown, both far 
upstream ($S_\infty$) and at the plane of symmetry ($S_0$). The geometry is shown for the 
cases of a spherical and a cylindrical obstacle as indicated.}
\label{fig:schematicRepresentation}
\end{figure}

A schematic view of the problem under investigation is depicted in figure \ref{fig:schematicRepresentation}. A 
suspended sphere of radius $a$ moves towards the obstacle parallel to one of the Cartesian coordinate axes, 
say the $x$-axis, from $x\to-\infty$. An obstacle of radius $b$ (spherical or cylindrical) is held fixed with its 
centre at the origin of coordinates. The centre-to-centre separation is given by the radial coordinate $r$, and 
the corresponding surface-to-surface separation, nondimensionalised by the mean of the two radii is 
$\xi = 2\left(r-a-b\right)/\left(a+b\right)$. The minimum (dimensionless) separation reached between the surfaces, 
$\xi_{min}$, occurs when the particle crosses the plane of symmetry perpendicular to the $x$-axis. We refer to the 
initial perpendicular distance between the line of motion and the $x$-axis as the {\it incoming impact parameter} 
and denote it by $b_{in}$ (it corresponds to the far upstream $y$-coordinate of the particle, as $x\to -\infty$). 
The motion is assumed to be caused by a uniform vector field $\boldsymbol{\mathcal{F}}$ acting on the sphere, 
which can be a constant body force ($\boldsymbol{\mathcal{F}}\equiv\mathbf{F}$, say, gravity) or 
a uniform incoming velocity field far from the obstacle 
($\boldsymbol{\mathcal{F}}\equiv 6\upi\mu a \mathbf{v^\infty}$). In either case, the sphere is torque free. 

We note that in the case of a spherical obstacle, the symmetry of the problem results in the
planar motion of the suspended particle. The centre of the particle moves in the plane 
formed by $\boldsymbol{\mathcal{F}}$ and the radial position vector at any time,
and the problem is axisymmetric (the unit vector in the radial direction is indicated by 
$\mathbf{d}$ as shown in figure \ref{fig:schematicRepresentation}; the plane of motion is the $xy$-
plane in the figure). In the case of a cylindrical obstacle, the motion of the particle parallel
to the axis of the cylinder is decoupled from that perpendicular to the axis, due to translational
symmetry. Here, we only consider the planar motion that occurs in the absence of a velocity component along the
axis of the cylinder. Further, since the driving field $\boldsymbol{\mathcal{F}}$ points along the 
positive $x$-axis, all trajectories in the problem are open, extending to infinity in both directions.

\subsection{Mobility functions for the problem}\label{subsec:mobFuncIntro}

In Stokes flows, the velocity of the suspended particle $\mathbf{U}$ is linear in the driving field 
$\boldsymbol{\mathcal{F}}$, such that $\mathbf{U}=\mathsfbi{M}~\boldsymbol{\cdot~\mathcal{F}}$ for an 
appropriate mobility tensor $\mathsfbi{M}$ that depends on the geometry of the particle-obstacle system. 
Moreover, since the particle motion is contained in the $xy$-plane, the velocity can be decomposed into two 
mutually orthogonal components in that plane; a radial component along $\mathbf{d}$ and a tangential 
component perpendicular to it \citep{batchelor1982, davis1992}. Then, given the symmetries of the 
system, the particle velocity can be represented as,

\begin{equation}
\mathbf{U} = (A(r) \mathbf{dd} + B(r) (\boldsymbol{\delta}-\mathbf{dd}))\cdot\boldsymbol{\mathcal{F}},
\label{eqn:velocityRepresentation}
\end{equation}

where $\boldsymbol{\delta}$ is the identity dyadic, and $A$ and $B$ are scalar functions of the 
separation $r$ (or equivalently, $\xi$), and also depend on the aspect ratio $\beta = b/a$.

It is well known that no analytical expressions are available for the radial ($A$) and tangential ($B$) 
mobility functions that are valid throughout the entire range of separations ($r\in\left(a+b,\infty\right)$ 
or $\xi\in\left(0,\infty\right)$). Instead, it is a common practice to derive expressions in two limiting 
cases, the near-field ($\xi\ll 1$) and the far-field ($\xi\gg 1$) limits \citep{jeffrey1984, kim1991}, and 
use some matching or interpolation procedure for intermediate separations. Lubrication theory and multipole expansions 
(equivalently, Lamb's general solution, method of reflections) are typically employed to derive mobility 
functions in the near-field and far-field, respectively. Available results in these 
limiting cases are briefly discussed below, as it will be useful for subsequent derivations. 
A more detailed discussion of the mobility functions used in this work is presented in appendix \ref{appx:mobFunc}.

\subsection{Mobility functions near contact $(\xi\ll1)$}\label{subsec:LubrProblem}
For sufficiently small separations, lubrication theory yields the following functional forms for the 
leading order of the radial and tangential mobility functions:
\begin{align}
A(\xi) &\approx a_0\xi \label{eqn:mobFuncLubrA}, \\ 
B(\xi) &\approx \frac{b_1}{\ln(1/\xi)} \label{eqn:mobFuncLubrB}.
\end{align}
Expressions are available for the coefficients $a_0$ and $b_1$ above, for any generic pair of 
convex surfaces that can be approximated by a quadratic form \citep{cox1974, claeys1989}. 
In particular, \citet{kim1991} provide the following expressions for $a_0$ and $b_1$ for a pair of spheres:
\begin{align*}
a_0 &= \frac{k_0}{6\upi\mu a}\left[\frac{\beta}{2}\left(1+\frac{1}{\beta}\right)^3\right]\notag, \\
b_1 &= \frac{k_1}{6\upi\mu a}\left[2\left(1+\frac{1}{\beta}\right)^3\right]\notag ,
\end{align*}
where, the constants $k_i=1$ for the case of a constant force acting on the particle, but are functions of $\beta$ in 
the case of a uniform velocity driving the particle. They can be written in terms of the $O(1)$-terms of 
the respective resistance functions, using the notation due to \citet{kim1991}:
\begin{align*}
k_0 &= A^X_{11} + \frac{1}{2}\left(1+\beta\right)A^X_{12}\notag, \\
k_1 &= A^Y_{11} + \frac{1}{2}\left(1+\beta\right)A^Y_{12} + \frac{1}{6}\frac{\beta + 4}{\beta + 1}\left(B^Y_{11} + \frac{1}{4}\left(1+\beta\right)^2 B^Y_{12}\right)\notag .
\end{align*}
The numerical values of $k_i$ for various values of $\beta$ are tabulated in Appendix A.

\citet{nitsche1996} applies expressions derived by \citet{claeys1989} to the case of a sphere moving relative to a 
cylinder of the same radius, in the case of a constant force driving the particle. We apply the same 
expressions to arbitrary ratios of radii to obtain:
\begin{align*}
a_0 &= \frac{1}{6\upi\mu a}\left[\frac{1+2\beta}{4}\left(1+\frac{1}{\beta}\right)^{3/2}\right],\\
b_1 &= \frac{1}{6\upi\mu a}\left[3\left(1+\frac{1}{\beta}\right)^{1/2}\right].
\end{align*}

\subsection{Mobility functions at large separations $(\xi\gg1)$}\label{subsec:mobFuncLargeXi}
In the case of a constant force driving the particle past a spherical obstacle, 
we use the far-field expansions of the mobility functions provided by \citet{kim1991} 
to obtain the following expressions for the functions $A(r)$ and $B(r)$ in powers of $(a/r)$:
\begin{align*}
A(r)  &=  \frac{1}{6\upi\mu a} \left[1-\frac{9}{4} \beta \left(\frac{a}{r}\right)^2+\left(\frac{3 \beta }{2}-\frac{9 \beta ^3}{4}\right) \left(\frac{a}{r}\right)^4 + O\left(\frac{1}{r^6}\right)\right],\\
B(r)  &=  \frac{1}{6\upi\mu a} \left[1-\frac{9}{16} \beta \left(\frac{a}{r}\right)^2-\frac{3}{8} \left(\beta +3 \beta ^3\right) \left(\frac{a}{r}\right)^4 + O\left(\frac{1}{r^6}\right)\right].
\end{align*}
Analogously, in the case of a uniform velocity field and a spherical obstacle we obtain,
\begin{align*}
A(r)  &= \frac{1}{6\upi\mu a} \left[1-\frac{3 \beta }{2} \left(\frac{a}{r}\right) + \frac{1}{2} \left(\beta + \beta ^3\right) \left(\frac{a}{r}\right)^3 + O\left(\frac{1}{r^5}\right)\right],  \\
B(r)  &= \frac{1}{6\upi\mu a} \left[1-\frac{3 \beta}{4} \left(\frac{a}{r}\right) - \frac{1}{4} \left(\beta + \beta^3\right) \left(\frac{a}{r}\right)^3 + O\left(\frac{1}{r^5}\right)\right].
\end{align*}
Note that, the leading order terms in the limit $a \ll b$ describe the streamlines of a uniform flow past a fixed 
spherical obstacle.

Finally, in the case of a constant force acting on a sphere moving past a cylindrical obstacle 
we use the following empirical far-field forms for radial and tangential mobility functions proposed by 
\citet{nitsche1996}:
\begin{align}
A(r)&=\frac{1}{6\upi\mu a}\left(1-\frac{117\upi}{128}\frac{a/r}{\ln(r/a\beta)+2.39}\right) \label{eqn:nitscheFFA}, \\
B(r)&=\frac{1}{6\upi\mu a}\left(1-\frac{48\upi}{128}\frac{a/r}{\ln(r/a\beta)+3.39}\right) \label{eqn:nitscheFFB}.
\end{align}
Here, we do not consider the case of a uniform velocity field driving a particle past an unbounded cylinder, since 
in this case, the description of the far-field motion in terms of mobility functions is not possible due to Stokes' 
paradox \citep{happel1965}.
In practice, however, the motion of the particle far from the obstacle would probably be dictated by the 
physical boundaries of the system (e.g. the walls of the microfluidic device). A similar `screening' argument 
will become relevant when we discuss the far-field asymptotic probability distribution in \S\ref{subsec:largeXiPDF}.
In order to understand the role of confinement, we shall consider the motion of a particle past a cylindrical obstacle between
two parallel walls forming a channel of width $2\ell_0$. We shall assume that the particle moves in the mid-plane,
driven by a constant force and thus the far-field mobility is dictated by the hydrodynamic interaction with the channel walls
\citep{happel1965}:
\[M_{wall}(r) = \frac{1}{6\pi\mu a}\left[1-1.004\left(\frac{a}{\ell_0}\right)+0.418\left(\frac{a}{\ell_0}\right)^3+0.21\left(\frac{a}{\ell_0}\right)^4-0.169\left(\frac{a}{\ell_0}\right)^5\right]\]
As explained in appendix A, at intermediate separations we interpolate between this far-field mobility and the 
mobility of a sphere in the vicinity of an infinite cylinder. The latter is obtained by an interpolation between 
(\ref{eqn:nitscheFFA}) (or (\ref{eqn:nitscheFFB})) and the lubrication regime discussed in \S\ref{subsec:LubrProblem} 
\citep{nitsche1996}.

\section{Steady-state distribution of particles around a fixed obstacle}\label{sec:steadyStateDistrib}

From an Eulerian point-of-view, we are interested in the steady-state distribution of 
particles around a fixed obstacle. The basic assumption in this investigation is 
that the particles are uniformly distributed far away from the obstacle, leading to a 
uniform flux of incoming particles at infinity.
The probability density of finding a particle at a given position $\mathbf{r}$ from 
the obstacle centre (i.e., the origin) is then analogous to the pair distribution function 
studied in dilute suspensions \citep{batchelor1972a, batchelor1982, davis1992,almog1997}. 
Thus, we talk of the pair distribution function $p(\mathbf{r},t)$, 
referring to particle-obstacle pairs, with the fixed obstacle located at the origin 
and the particle at position $\mathbf{r}$. Further, $p({\bf r},t)$ is the normalised 
distribution function, such that the actual probability density is given by $n_p p({\bf r},t)$, 
where $n_p$ is the uniform number density of particles far from the obstacle.
\begin{figure}
\begin{center}
\includegraphics[scale=0.5]{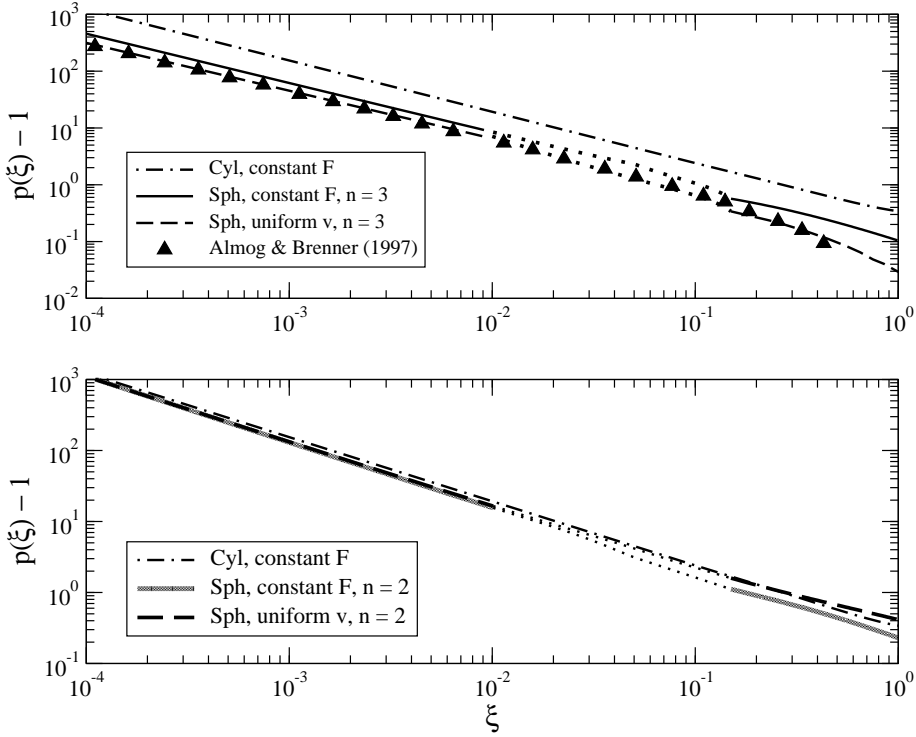}
\end{center}
\caption{Excess probability distribution ($p(\xi)-1$) as a function of the dimensionless 
surface-to-surface separation, for radius-ratio $\beta = 1$. (top) Comparison between both 
flow cases for a spherical obstacle with $n=3$ and a cylindrical obstacle. (bottom) 
Comparison between both flow cases for a spherical obstacle with $n=2$ and a cylindrical 
obstacle.}
\label{fig:p-1SphCylCF-SphSphCF-SphSphCV}
\end{figure}

Following \citet{batchelor1972a}, we start with the conservation equation for the number 
of particles, which in terms of the pair distribution function can be written as follows:
\begin{equation}
\frac{dp}{dt}=\frac{\partial p}{\partial t} + {\mathbf U}\cdot\nabla p = -p \nabla\cdot {\mathbf U}.
\label{eqn:FokkerPlank}
\end{equation}
Then, we write the divergence of the particle velocity in terms of its radial component
using (\ref{eqn:velocityRepresentation}),
\begin{equation}
\nabla\cdot{\mathbf U} = \left[\frac{(n-1)(A-B)}{rA}+\frac{1}{A}\frac{dA}{dr}\right]U_r,
\label{eqn:velocityDivergence}
\end{equation}
where $n$ equals $2$ or $3$ depending on the dimensionality of the problem (see the discussion below). 
Substituting this expression into (\ref{eqn:FokkerPlank}) we obtain,
\begin{equation}
\frac{d p}{d t} = - p U_r \left[\frac{(n-1)(A-B)}{rA}+\frac{1}{A}\frac{dA}{dr}\right] = p U_r \frac{1}{q}\frac{dq}{dr}
=p \frac{dr}{dt} \frac{1}{q} \frac{dq}{dr} = \frac{p}{q}\frac{dq}{dt} ,
\label{eqn:defineFuncQ}
\end{equation}
where, following \citet{batchelor1972a} we introduced a function $q(r)$ defined 
such that $p(\mathbf{r},t)/q(r)$ is a constant of motion (its material derivative 
is zero). This result is valid for both types of obstacles and both driving fields, 
with $q(r)$ determined by the mobility functions specific to each case. Further, 
$p(\mathbf{r},t)/q(r)$ is constant under both steady as well as unsteady conditions and, 
as a result, the distribution of particles at a position 
$\mathbf{r}$ and time $t$ can be determined from a previous position of the same 
material point at a prior time.
In addition, in steady state $p(\mathbf{r})/q(r)$ is constant along trajectories,
which implies a stationary, radially symmetric distribution of particles
(i.e., $p(\mathbf{r}) \equiv p(r)$), a surprising result, but one that is analogous to 
that obtained by \citet{batchelor1972a}.
An alternative way to arrive at the same result is to propose a function $q(r)$ such that 
the field $q \mathbf{U}$ becomes solenoidal and therefore $p/q$ is constant along particle 
trajectories. We can obtain a solenoidal field by multiplying the velocity with a function 
$q$ of the magnitude $r$ only, which indicates an isotropic distribution of particles in 
steady state. Finally, solving for $q(r)$ 
and imposing the asymptotic limit $p(r\to\infty) \to 1$, we obtain
a general expression for the pair distribution function in terms of the mobility functions A and B, 
\begin{equation}
p(r)=\frac{A^\infty}{A(r)}\exp\left\{\displaystyle\int_r^\infty \frac{(n-1)(A-B)}{r A} dr \right\},
\label{eqn:pairDistributionFunc}
\end{equation}
where $A^\infty= 1/(6\pi\mu a)$ is the asymptotic value of the radial mobility function for 
$r\to\infty$.
The precise effect of geometry and driving field is captured by the hydrodynamic mobility 
functions $A$ and $B$, as well as the effective {\em dimensionality} $n$ of the problem. 
An infinitely long cylindrical obstacle naturally imposes a two dimensional geometry ($n=2$). 
However, there are two independent possibilities in the case of a spherical obstacle.
First, the fully 3-dimensional (3D) problem of a dilute suspension moving past a spherical obstacle 
corresponds to $n=3$.
The second problem corresponds to the case in which the suspended particles are 
restricted to move in the $xy$-plane, i.e., the plane of the paper in figure 
\ref{fig:schematicRepresentation}. This simplification corresponds to $n=2$, and is sometimes 
used to approximate the motion past a cylindrical fibre \citep{phillips1989, phillips1990}.

In figure \ref{fig:p-1SphCylCF-SphSphCF-SphSphCV}, we show the probability density 
function for an aspect ratio $\beta = 1$ and various possibilities in terms of driving field,
geometry of the obstacle, and the associated dimensionality. 
The top plot shows that the distribution of particles around a cylindrical obstacle is almost five-fold 
larger than that in the case of a spherical obstacle with $n=3$ for a given separation from the obstacle. 
The distribution is observed to be less sensitive to the effect of the driving field (spherical obstacle, 
$n=3$), with that for the particles driven by a constant force being higher than when they are 
driven by a uniform velocity field. The same plot also shows a good agreement 
between our calculations for a spherical obstacle (with $n=3$) and those of \citet{almog1997}, 
for the case of a particle driven by a uniform velocity field. Note that, \citet{almog1997} have 
analysed the equivalent case of a particle moving with a prescribed constant velocity in 
a suspension of neutrally buoyant spheres.
In the bottom plot we compare the probability distribution around a cylindrical obstacle with
that around a spherical obstacle when the incoming particles are restricted to move in the $xy$-plane ($n=2$). 
The distributions exhibit similar behaviour at small separations. However, we will show that 
the asymptotic scaling of the probability distributions in the limit of small separations is different for each 
geometry of the obstacle. We also note that all the distributions are in fact divergent 
at contact ($\xi \to 0$). We shall discuss the asymptotic behaviour and this divergence of the 
probability distributions in more detail in \S\ref{subsec:smallXiPDF}.
\begin{figure}
\begin{center}
\includegraphics[scale=0.5]{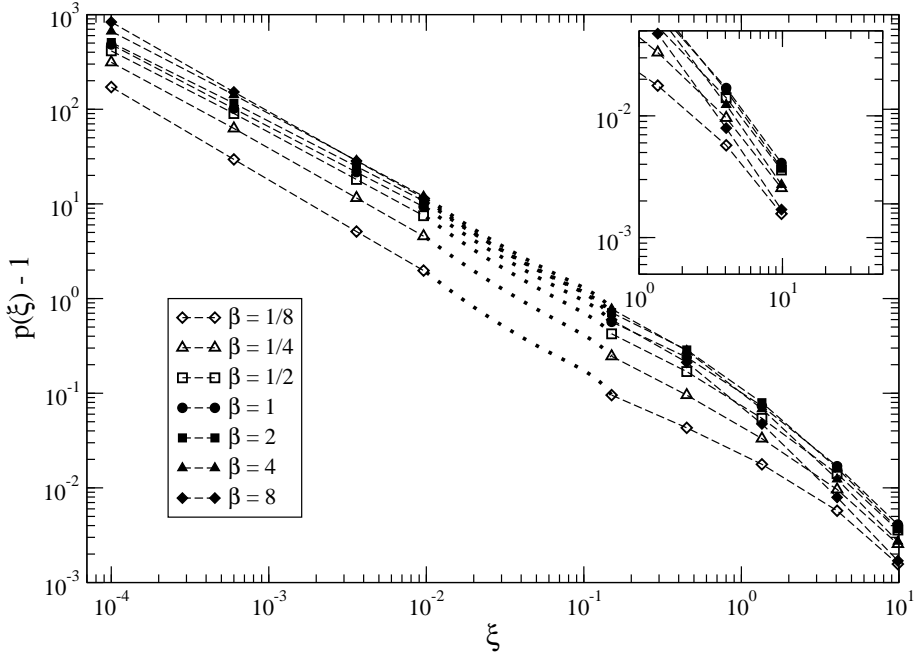}
\end{center}
\caption{Excess probability distribution ($p(\xi)-1$) as a function of dimensionless surface-to-surface 
separation, for different aspect ratios $\beta$, a constant force and a spherical obstacle ($n=3$). 
Inset: Enlarged view for large values of $\xi$, depicting symmetric behaviour about $\beta=1$ for a fixed 
value of $\xi$, i.e., filled and open symbols almost overlap.}
\label{fig:p-1AllBetaSphSphCF}
\end{figure}

In figure \ref{fig:p-1AllBetaSphSphCF}, we compare the probability distribution function for different 
aspect ratios between the suspended particle and a spherical obstacle at the centre ($n=3$) when the 
particle is driven by a constant force. 
The distribution of particles increases monotonically with the aspect ratio $\beta$ for relatively 
small separations, $\xi \lesssim1$. However, for large separations the probability distribution function 
exhibits a symmetric scaling behaviour with respect to the aspect ratio, in that the same leading order 
expressions are obtained for $\beta$ and $\beta^{-1}$, as discussed in \S\ref{subsec:largeXiPDF} below.

As mentioned before, the mobility functions used to calculate the probability distribution plotted in 
figures \ref{fig:p-1SphCylCF-SphSphCF-SphSphCV} and \ref{fig:p-1AllBetaSphSphCF} in the case of a spherical 
obstacle were obtained from the literature for the limiting cases of small and large separations. For intermediate 
separations, a transition is made between the two limits using interpolation. This interpolation 
region is depicted by dotted lines in figures \ref{fig:p-1SphCylCF-SphSphCF-SphSphCV} and 
\ref{fig:p-1AllBetaSphSphCF}. For the cylindrical obstacle, we have used interpolated functions empirically 
computed by \citet{nitsche1996} over the entire range of separations (see Appendix \ref{appx:mobFunc}).

\subsection{Asymptotic behaviour of the probability distribution near contact}\label{subsec:smallXiPDF}
In order to investigate the asymptotic behaviour of the probability distribution near contact, we first 
rewrite (\ref{eqn:pairDistributionFunc}) in terms of the dimensionless surface-to-surface separation $\xi$,
\begin{equation}
p(\xi)=\frac{A^\infty}{A(\xi)}\exp\left\{(n-1) H(\xi)\right\},
\label{eqn:pairDistributionFuncNonDim}
\end{equation}
where,
\begin{equation}
H(\xi) = \displaystyle\int_\xi^\infty \frac{(A-B)}{2\left(1+\frac{\tilde{\xi}}{2}\right) A} d\tilde{\xi}.
\label{eqn:definitionOfH}
\end{equation}

\begin{figure}
\begin{center}
\includegraphics[scale=0.5]{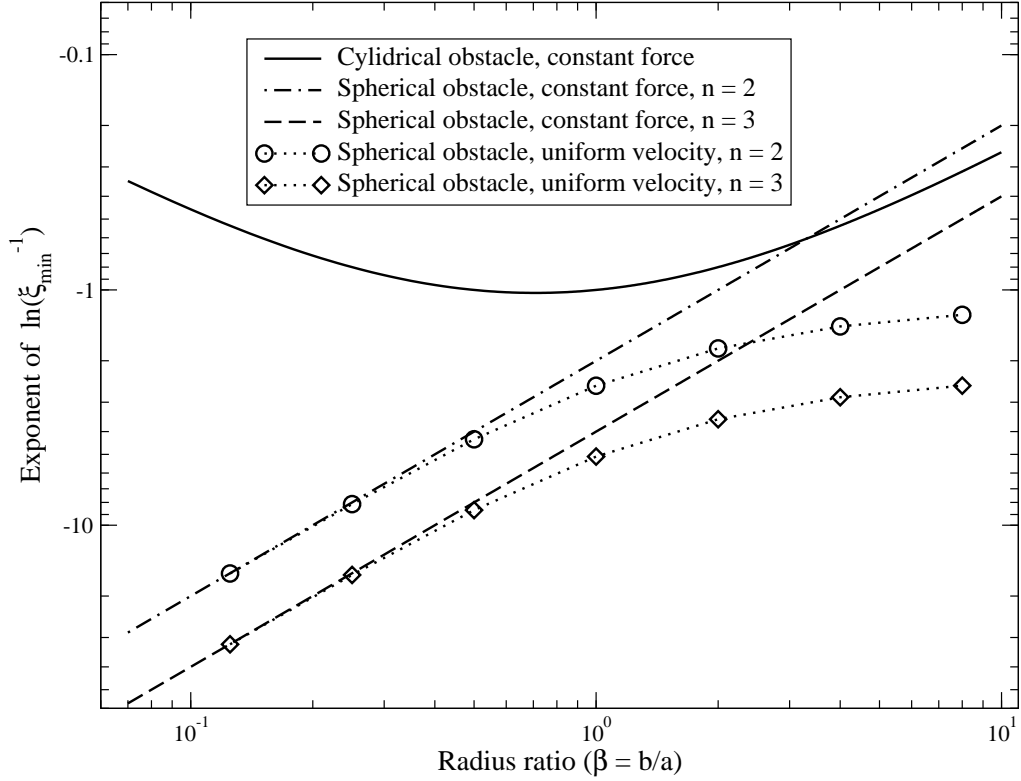}
\end{center}
\caption{Exponents of $\ln(\xi^{-1}_{min})$ from (\ref{eqn:approxPairProbability}), (\ref{eqn:scalingEquation}) 
and (\ref{eqn:improvedScaling}) as a function of $\beta$ for both cylindrical and spherical obstacles. The exponent 
for the case of a uniform velocity driving a particle past a sphere ($\Diamond$ and $\bigcirc$) is evaluated at 
discrete values of $\beta = 1/8,~1/4,~1/2,~1,~2,~4,~8$, as explained in the text.}
\label{fig:exponentsSmallXiFuncBeta}
\end{figure}
For small separations, we can split the domain of integration for $H(\xi)$ at some arbitrary value $\xi_0 \ll 1$ 
such that the expressions for the mobility functions $A$ and $B$ can be approximated using 
lubrication theory in the near-field part of the integral, where $\xi \ll 1$. From this lubrication region 
we obtain the leading order term for the probability distribution of particles at small separations,
\begin{equation}
p(\xi)\approx p_0 \xi^{-1}(\ln(\xi^{-1}))^{-(n-1)\alpha}~~~\ldots~\alpha = \frac{b_1}{2 a_0}~,
\label{eqn:approxPairProbability}
\end{equation}
where $p_0$ is a constant that depends on the geometry and driving field, and from 
\S\ref{subsec:LubrProblem}, $\alpha = 2k_1/k_0\beta$ for the case of a spherical obstacle, while 
$\alpha = 6\beta(1+2\beta)^{-1}(1+\beta)^{-1}$ for the case of a constant force driving the particle 
past a cylindrical obstacle.
The divergence of the probability distribution at small separations is dominated by the
$\xi^{-1}$ term, which explains the apparent linear behaviour observed in figure 
\ref{fig:p-1SphCylCF-SphSphCF-SphSphCV} and the weak dependence on the aspect ratio observed in figure
\ref{fig:p-1AllBetaSphSphCF}. 
The asymptotic expression (\ref{eqn:approxPairProbability}) is similar to the results 
obtained by \citet{batchelor1982} and \citet{khair2006a}. They obtain an expression 
of the form $p(\xi)\sim\xi^{-\ell}(\ln(\xi^{-1}))^{-m}$, wherein $\ell (\ne1)$ arises due 
to the presence of $O(\xi^0)$ terms in their tangential mobility functions. These terms 
are exactly zero when the obstacle is fixed, leading to $\ell = 1$, as obtained in 
(\ref{eqn:approxPairProbability}). Figure \ref{fig:exponentsSmallXiFuncBeta} shows the exponent 
$-(n-1)\alpha$ in the equation above, as a function of $\beta$ for both types of obstacles. 
As mentioned in \S\ref{subsec:LubrProblem}, $k_1 = k_0 = 1$ for the case of a constant force 
driving the particle past a spherical obstacle. This leads to $\alpha = 2/\beta$, corresponding
to the two straight lines in the log-log plot presented in the figure. Further, for the case of a uniform velocity 
driving the particle, $k_0$ and $k_1$ can be computed using the numerical values of the $O(1)$ terms in the 
corresponding resistance functions, which are available at discrete values of the aspect ratio 
\citep{jeffrey1984, kim1991}. This leads to a numerical evaluation of the exponent at these values, as shown 
in the figure (see appendix A for a table of values of $k_0$ and $k_1$). We note that for very small values 
of the aspect ratio$\beta$ -- i.e., when the obstacle is very small compared to the particle -- we get 
$k_0\approx k_1\approx1$, and the exponent for the case of a spherical obstacle becomes equal for both 
types of driving fields. This is consistent with a constant drag acting on the suspended particle outside 
the lubrication region.

Finally, we note the presence of a diffusive boundary layer around the obstacle. A local P\'{e}clet 
number for the particle, $Pe_L$, can be written as the ratio between the radially inward convective 
flux ($j_C = n_P p(\xi)U_r$) and the radially outward diffusive flux ($j_D = - n_P D dp/dr$). At small 
separations we can use the asymptotic behaviour of $p(\xi)$ to obtain,
\begin{align*}
Pe_L &= \frac{j_C}{j_D} = -\frac{n_P p U_r}{n_P D dp/dr} = -\frac{p(\xi)}{dp/d\xi}\cdot\frac{(a+b)\mathcal{F}}{2kT} 
\sim~ \xi Pe_0,
\end{align*}
where, $Pe_0=(a+b)\mathcal{F} / {2kT}$ is the system P\'{e}clet number. 
This indicates the existence of a boundary layer around the obstacle for $\xi\lesssim 1/Pe_0$, 
within which the diffusive mode of transport dominates over the convective one.
Thus, the expressions derived above for the probability distribution, under the deterministic assumption,
are not valid within this boundary layer.
\citet{batchelor1982} highlights this issue for sedimenting suspensions.
\citet{nitsche1996} discusses the presence of such a boundary layer in the context of the 
diffusion of a spherical particle close to a cylindrical fibre of comparable size and 
\citet{khair2006a} have quantified this region in the context of micro-rheology.

\subsection{Asymptotic behaviour of the probability distribution at large distances}\label{subsec:largeXiPDF}
In order to obtain the leading order behaviour of the probability distribution far from the obstacle, we
use the expressions for the far-field mobility functions discussed in \S\ref{subsec:mobFuncLargeXi}. 
Substituting these far-field expressions for the mobility functions $A$ and $B$ 
in (\ref{eqn:pairDistributionFunc}), we obtain the asymptotic probability distribution for 
$\xi\gg1$,
\begin{equation}
p^F_{Sph}(r) \sim 1+\frac{45}{32} \beta  \left(\frac{a}{r}\right)^2 + O\left[\left(\frac{a}{r}\right)^4\right],
\label{eqn:largeSeparationsScalingn2}
\end{equation}
\begin{equation}
p^v_{Sph}(r) \sim 1+\frac{3}{4}\beta\frac{a}{r}+\frac{27}{32} \beta ^2 \left(\frac{a}{r}\right)^2 + O\left[\left(\frac{a}{r}\right)^3\right],
\label{eqn:largeSeparationScalingCVn2}
\end{equation}
\begin{equation}
p^F_{Cyl}(r) \sim 1 + \frac{1.18 (a/r)}{\left(3.39 + \ln\left[\frac{r}{a\beta}\right]\right)} + O\left[\left(\frac{a}{r}\right)^2\right],
\label{eqn:largeSeparationScalingCylCF}
\end{equation}
corresponding to a spherical obstacle ($n=2$) and constant driving force, 
a spherical obstacle ($n=2$) and a uniform velocity field driving 
the particle, and a cylindrical obstacle and a constant driving force, 
respectively. 

Further, for a spherical obstacle with $n=3$, the following expressions for the asymptotic distribution 
are obtained:
\begin{equation}
p^F_{Sph}(r) \sim 1+\frac{9}{16} \beta  \left(\frac{a}{r}\right)^2 + O\left[\left(\frac{a}{r}\right)^4\right],
\label{eqn:largeSeparationsScalingn3}
\end{equation}
\begin{equation}
p^v_{Sph}(r) \sim 1+\frac{27}{8} \beta^2  \left(\frac{a}{r}\right)^5 + O\left[\left(\frac{a}{r}\right)^6\right],
\label{eqn:largeSeparationsScalingCVn3}
\end{equation}
Note that both (\ref{eqn:largeSeparationsScalingn2}) and (\ref{eqn:largeSeparationsScalingn3})  
are symmetric about $\beta = 1$ for a given (albeit large) separation $\xi$. 
This can be observed in figure \ref{fig:p-1AllBetaSphSphCF} for $\xi \sim 10$ (see inset).
When the particle and the obstacle can be treated as point forces separated far apart, the 
expressions the for mobility functions (given in \S\ref{subsec:mobFuncLargeXi}) become symmetric 
in their radii $a$ and $b$, which leads to the symmetry shown above. In contrast, the mobility 
functions and the corresponding probability distribution remain asymmetric in the case of a 
uniform velocity field driving the particles.

Analogous expansions in the case of suspension flows are discussed by \citet{batchelor1982} and by 
\cite{almog1997} in the context of micro-rheology. However, there is an important difference with these cases.
The expressions for $n=2$, (\ref{eqn:largeSeparationsScalingn2}), (\ref{eqn:largeSeparationScalingCVn2}) and 
(\ref{eqn:largeSeparationScalingCylCF}) (for $n=3$, (\ref{eqn:largeSeparationsScalingn3})) indicate that the 
integral $\int_{(a+b)}^R[p(r)-1]2\pi rdr$ (correspondingly, $\int_{(a+b)}^R[p(r)-1]4\pi r^2dr$) diverges for $R\to\infty$. 
Physically, this integral indicates a divergence in the total excess number of particles in the suspension at steady 
state, induced by the long-range nature of the hydrodynamic interactions with the fixed obstacle. As we 
mentioned earlier in \S\ref{subsec:mobFuncLargeXi}, in the case of microfluidics either the channel walls 
and/or the presence of other suspended particles would effectively screen such long-range interactions, thus 
leading to convergent integrals. On the other hand, the asymptotic particle distribution around a spherical obstacle 
for the case of a uniform velocity field and $n=3$ given by (\ref{eqn:largeSeparationsScalingCVn3}), 
yields a finite excess number of particles. We attribute this to the 
rapidly decaying divergence of the particle velocity in this case, given by (\ref{eqn:velocityDivergence}, $n=3$),
\[\nabla\cdot{\mathbf U} \propto \ln[p(r)] \sim \frac{27}{8}\beta^2\left(\frac{a}{r}\right)^5~~~\ldots~~~r\gg1.\]

\section{Minimum surface-to-surface separation}\label{sec:trajectoryAnalysis}
We now turn to the dependence of the dimensionless minimum surface-to-surface separation, 
$\xi_{min}$ (equivalently, the corresponding centre-to-centre separation $r_{min}$), on 
the incoming impact parameter $b_{in}$, following both Eulerian and 
Lagrangian approaches. The Eulerian approach entails invoking the conservation of the number 
of particles entering and exiting the region defined by the revolution of the trajectory 
of a particle around the $x$-axis (the region created by the translation of a particle 
trajectory along the axis of the cylinder in the case of a cylindrical obstacle). The 
Lagrangian approach involves the explicit calculation of the trajectory of an individual 
particle, followed by the evaluation of the position of the particle as it crosses the plane 
$x=0$.

\begin{figure}
\begin{center}
\includegraphics[scale=0.5]{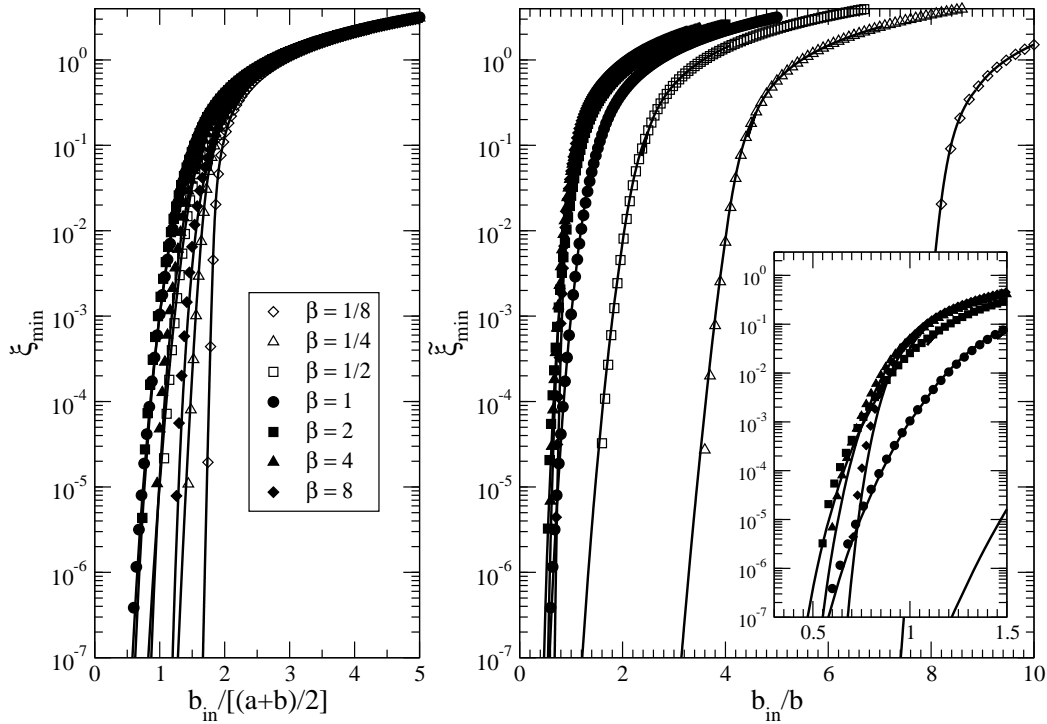}
\end{center}
\caption{Minimum separation $\xi_{min}$ versus incoming impact parameter $b_{in}$. Comparison 
between particle-particle simulations \citep{frechette2009} and the analytical result given in 
(\ref{eqn:governingRelation}): different symbols depict the simulation data as 
described in the legend, while the solid lines represent (\ref{eqn:governingRelation}). 
(left) Average radius, $(a+b)/2$, is used to nondimensionalize the axes, (right) radius of 
the obstacle is used to nondimensionalize the axes, $\tilde{\xi} = [(1+\beta)/2\beta]\xi$, 
(inset) Enlarged view of the interval $(0.5,1.5)$ of $\tilde{b}_{in}$}
\label{fig:betaAllCompareJFM}
\end{figure}

\begin{figure}
\begin{center}
\includegraphics[scale=0.5]{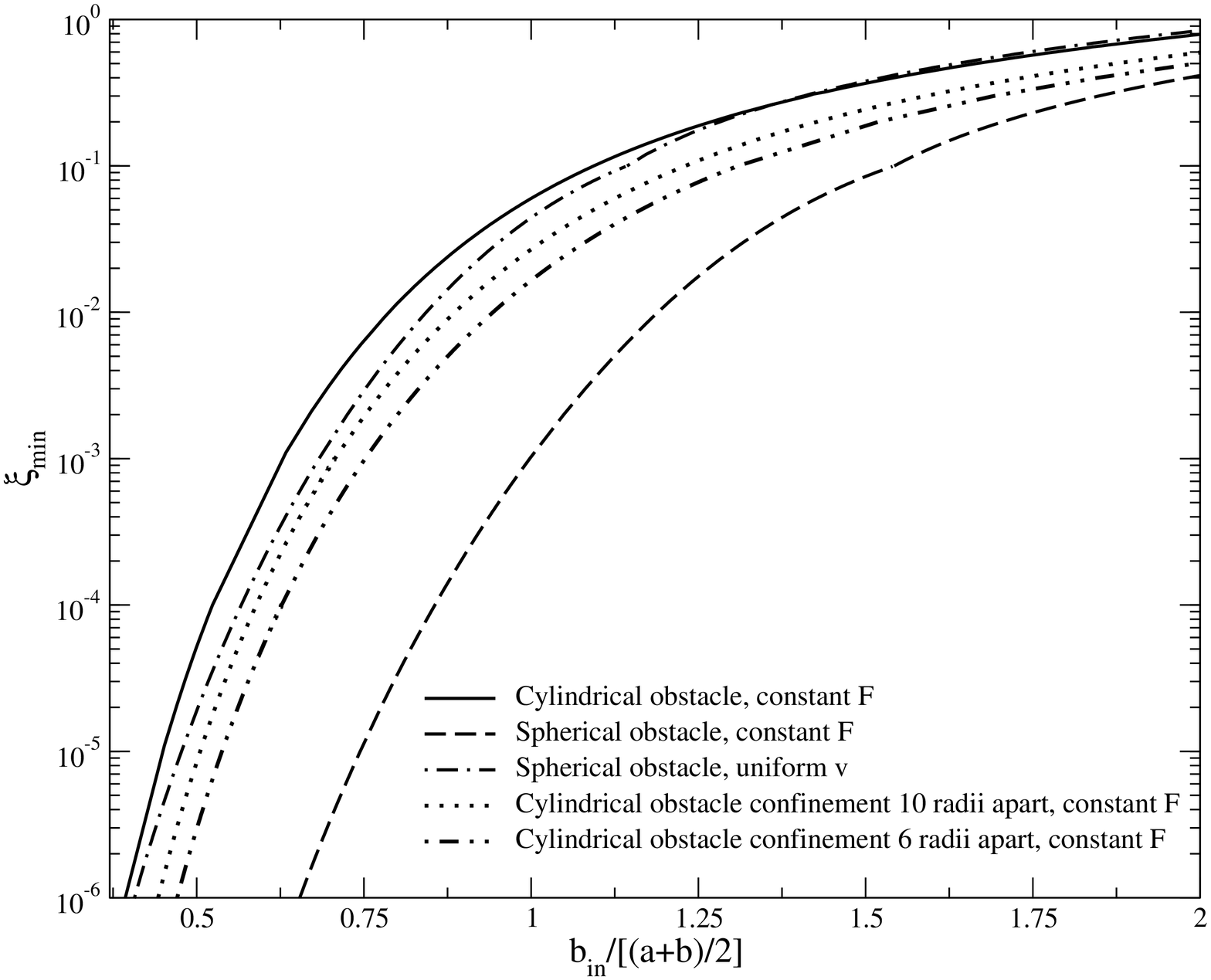}
\end{center}
\caption{Minimum separation $\xi_{min}$ versus incoming impact parameter $b_{in}$ for different 
geometries and flow-fields when the particle and the obstacle are of the same radius. The incoming 
impact parameter is made dimensionless using the average radius, $(a+b)/2$.
As explained in the text, for a cylindrical obstacle, the minimum separation decreases for a given 
incoming impact parameter in case of a confined system.}
\label{fig:compareAllFlowGeomBeta1}
\end{figure}
\subsection{Eulerian approach: Conservation of the number of particles}\label{subsec:trajAnalEuler}
We consider the situation depicted in the bottom half of figure \ref{fig:schematicRepresentation}. 
On the left, it shows the volume of revolution obtained by revolving around the axis of motion ($x$-axis)
the trajectory followed by a spherical particle moving past a spherical obstacle. 
The conservation of particles inside this volume of revolution takes a simple form, 
due to the fact that the flux of particles across the surface of revolution is zero; since trajectories do not 
cross in Stokes flow. (The construction of such a volume of revolution
corresponds to the fully 3D case with $n=3$. The 2D case with $n=2$ can be treated in a manner completely analogous
 to that of a cylindrical obstacle.) Then, we calculate the flux at two cross sections that are perpendicular 
to the $x$-axis, and over which the velocity is exactly normal to the surface. These two surfaces are depicted 
in figure \ref{fig:schematicRepresentation}: one is the cross section far upstream, indicated by $S_\infty$, and 
the otheris the annular region corresponding to the cross section at $x=0$, indicated by $S_0$ in the figure.
(Note the excluded volume of radius $(a+b)$). On the bottom right corner of the figure, we show the 
analogous case of a cylindrical obstacle, in which the flux is considered per unit length in 
the $z$-direction. In both cases the local flux of particles is given by  $p(\mathbf{r})\mathbf{U}$. 
Therefore, the conservation of particles in terms of surface integrals takes the form, 
\begin{equation}
\displaystyle\int_{S_\infty} \left.p\mathbf{U}\right|_{x\rightarrow -\infty}\cdot d\mathbf{S} = 
\displaystyle\int_{S_\infty} p_\infty \, U_\infty \, dS = 
\displaystyle\int_{S_0} p_0 \, U_0 \, dS =
\displaystyle\int_{S_0} \left.p\mathbf{U}\right|_{x=0} \cdot d\mathbf{S},
\label{eqn:fluxOfParticles}
\end{equation}
where $U_\infty$ and $U_0$ are the magnitude of the particle velocity far upstream and at $x=0$, respectively.
The integration over $S_\infty$ is simple, the probability distribution tends to unity and the
velocity is uniform. Specifically, the velocity is asymptotically radial and
equal to $A^\infty\mathcal{F}$, that is $F/6\pi \mu a$ or $v^{\infty}$, depending on the driving field.
The flux is, therefore, $S_{\infty} A^\infty \mathcal{F}$, with $S_\infty = \pi^{(n-2)} b^{(n-1)}_{in}$. 
At the plane of symmetry, the velocity is tangential to the obstacle, $U_0=B(r)\mathcal{F}$. Then,  substituting in (\ref{eqn:fluxOfParticles}) and
performing suitable algebraic simplifications we obtain,
\begin{equation}
\frac{b_{in}^{n-1}}{n-1} = \displaystyle\int_{a+b}^{r_{min}}\frac{B(r)}{A(r)}r^{n-2}\exp\left\{\displaystyle\int_r^\infty\frac{(n-1)(A-B)}{\tilde{r} A}d\tilde{r}\right\}dr,
\label{eqn:unsimplifiedGoeverningRelation}
\end{equation}
where the limits of integration of the annular region are the excluded volume radius $(a+b)$ and the
radial position of the particle at $x=0$, which corresponds to the minimum radial distance
reached during the course of the motion, $r_{min}$. We can simplify the equation further, 
as shown in Appendix \ref{appx:simplificationOfGoverningRelation} to yield,
\begin{equation}
b_{in} = r_{min}\exp\left\{H\left(\xi_{min}\right)\right\},
\label{eqn:governingRelation}
\end{equation}
where $H$ is the function defined in (\ref{eqn:definitionOfH}).

\subsection{Lagrangian approach: Trajectory analysis}\label{subsec:trajAnalLagrange}

The differential equation for the particle trajectory in terms of $y$ and $r$ coordinates can be
integrated in a straightforward manner by separation of variables,
\begin{equation}
\frac{dy}{dr} = \frac{U_y}{U_r} = \frac{U_r\sin\theta + U_\theta \cos\theta}{U_r} = \sin\theta -\frac{B(r)\mathcal{F}\sin\theta}{A(r)\mathcal{F}\cos\theta}\cos\theta = y\left[\frac{1}{r}-\frac{B(r)}{r~A(r)}\right].
\label{eqn:particleTrajectoryLagrangian}
\end{equation}
Thus, upon integration between $(y=b_{in},r\to\infty)$ and $(y,r)$ we get,
\begin{equation}
\nonumber
\ln\left(\frac{b_{in}}{y}\right) = \displaystyle\int_r^\infty\frac{A(r)-B(r)}{r A(r)} dr = H(\xi) 
\Rightarrow b_{in} = y \exp\{H(\xi)\}.
\end{equation}
Evaluating at $x=0$ (where $y=r_{min}$ and $\xi=\xi_{min}$), we retrieve (\ref{eqn:governingRelation}).

The expression (\ref{eqn:governingRelation}) explicitly relates the minimum separation between 
particle and obstacle surfaces with the incoming impact parameter, and is a general result 
applicable to systems in which the velocity can be decomposed as in (\ref{eqn:velocityRepresentation}). 
It is interesting to note that it has the same \emph{form} 
for the relation between $b_{in}$ and $r_{min}$ (or, $\xi_{min}$) for two and three dimensional 
geometries, which can be explained by the fact that the motion is planar in both cases. 
Again, the entire information about the driving field and the geometry of the problem is 
implicit in the mobility functions $A(r)$ and $B(r)$.

In figure \ref{fig:betaAllCompareJFM}, we compare the minimum separation obtained using 
(\ref{eqn:governingRelation}) with that obtained from the numerical integration of 
the trajectory followed by a spherical particle moving around a fixed spherical obstacle \citep{frechette2009}. 
Excellent agreement is observed for all the aspect ratios considered, $\beta = 0.125, 0.25, 0.5, 1, 2, 4, 8$. 
Note that the same mobility functions were used in both cases, hence the agreement is independent 
of the accuracy of the mobility functions themselves. 
In order to illustrate (\ref{eqn:governingRelation}) in a practical manner, we also present the minimum 
separation for particles of different size when the size of the obstacle is fixed (see right hand panel 
in figure \ref{fig:betaAllCompareJFM}). We observe that for a constant incoming impact parameter, 
$b_{in}/b\sim O(1)$, the minimum attained separation decreases with increasing particle radius. However, an 
asymptotic analysis at very small separations reveals that for a given (sufficiently small) incoming impact parameter, 
the minimum separation decreases monotonically with decreasing particle radius. This can be corroborated from the 
behaviour of the exponent involved in the asymptotic scaling of the incoming impact parameter (\ref{eqn:scalingEquation}) 
shown in figure \ref{fig:exponentsSmallXiFuncBeta} (see \S\ref{subsec:smallXi} below for a detailed discussion). The inset in 
the right hand panel of figure \ref{fig:betaAllCompareJFM}, in fact, portrays the imminent reversal of this trend 
in the form of the criss-crossing curves.
Figure \ref{fig:compareAllFlowGeomBeta1} shows $\xi_{min}$ as a function 
of $b_{in}$ for $\beta=1$ for five different cases. We can see that for a constant driving force and a given 
$b_{in}$ the particle gets closer to a spherical obstacle, compared to a cylindrical one. It also shows that 
a constant force drives the particle closer to a spherical obstacle than a uniform 
velocity field. Further, we show the case of an obstacle-particle pair confined between two parallel walls, 
when the obstacle is a cylinder and the walls are perpendicular to the axis of the cylinder. This case 
is the most relevant to model the motion of a particle past a cylindrical post in a microfluidic 
channel. As explained in \S\ref{subsec:mobFuncLargeXi} and appendix A, at intermediate separations 
we compute the mobility of the particle in this system by interpolating between the mobility of a sphere moving
along the mid-plane between the two walls \citep{happel1965} and that for a particle moving close to an infinite 
cylinder \citep{nitsche1996}. We observe that the 
presence of such confinement, as well as increasing the extent of the confinement from 10 particle radii to 6 
particle radii, decreases the attained minimum separation for a given incoming impact parameter. The underlying 
reason is that, in the far-field, the mobility is isotropic in the plane of motion and, therefore, the motion of 
the particle is not affected by the obstacle. The interaction with the obstacle becomes significant for 
particle-obstacle separations of the order of the channel width. Thus, for a narrower channel, the particle reaches 
closer to the obstacle before its mobility is affected by the presence of the obstacle. 
The minimum separation attained by the particle along its trajectory determines the relevance of short-ranged 
repulsive forces during its motion. Thus, in terms of the effect of 
confinement, the above observation implies that the particle will experience an earlier onset of non-hydrodynamic
effects upon confinement between two parallel walls.

\subsection{Asymptotic behaviour of minimum separation near contact, $\xi_{min}\to0$}\label{subsec:smallXi}
Here, we calculate the asymptotic behaviour of the relation (\ref{eqn:governingRelation}) in the limit of very small separations,
in a manner similar to that discussed in \S\ref{subsec:smallXiPDF} for the probability distribution. 
Specifically, using the lubrication approximation for the mobility functions in \S\ref{subsec:LubrProblem}, 
and substituting in (\ref{eqn:governingRelation}) we get:
\begin{equation}
\left(\frac{b_{in}}{(a+b)/2}\right)\approx \lambda_0 (\ln(1/\xi_{min}))^{-\alpha}~~~\ldots~\alpha = \frac{b_1}{2 a_0},
\label{eqn:scalingEquation}
\end{equation}
where $\lambda_0$ is a constant that incorporates the contributions from the far-field.
The behaviour of the exponent $(-\alpha)$ is depicted in figure 
\ref{fig:exponentsSmallXiFuncBeta}. As discussed earlier, we can see from the behaviour of the exponent
that for incoming imapct parameters small enough to follow (\ref{eqn:scalingEquation}), the minimum separation 
decreases monotonically with decreasing particle radius.

However, the asymptotic form of the tangential resistance function $1/B(r)$, derived using (\ref{eqn:mobFuncLubrB}),
is only logarithmically divergent, and valid only when extremely small separations are attained. 
Thus, for reasonably small values of $\xi_{min}$, 
albeit $\xi_{min}\ll1$, the expression (\ref{eqn:scalingEquation}) is not an accurate approximation. Therefore, in addition to 
the leading order of the scalar resistance functions from which $B(\xi)$ is obtained by matrix inversion, we also retain their $O(1)$ 
terms. Consequently, the tangential mobility function takes the form:
\begin{equation}
B(\xi) \approx \frac{1}{6\upi\mu a}\frac{s(\xi)(b_1 + b_2 s(\xi))}{1 + b_3 s(\xi) + b_4 s(\xi)^2}~~~\ldots~~~s(\xi) = \frac{1}{\ln(1/\xi)}~.
\label{eqn:newApproxB}
\end{equation} 
We continue to use (\ref{eqn:mobFuncLubrA}) as an approximation for the radial mobility function, so that $A(\xi)\approx a_0\xi$. 
This leads to the following approximate expression for the scaling:
\begin{equation}
\left(\frac{b_{in}}{(a+b)/2}\right)\approx \frac{\lambda_1(s(\xi_{min}))^\alpha}{\left(\gamma_1+s(\xi_{min})\right)^{\alpha-\gamma_0}\left(\gamma_2+s(\xi_{min})\right)^{\gamma_0}},
\label{eqn:improvedScaling}
\end{equation}
where, as before, the exponent $\alpha$ is as defined in (\ref{eqn:approxPairProbability}) and (\ref{eqn:scalingEquation}), 
and the constant $\lambda_1$ captures the far-field behaviour. 
The constants 
$\gamma_i$ are functions of the coefficients $b_i$ in the tangential mobility function (see appendix A for 
their tabulated values). In figure \ref{fig:scalingEquationCompare}, we compare the limiting behaviour given by 
(\ref{eqn:improvedScaling}), with $\lambda_1$ as the only fitting parameter, with the numerical evaluation of the 
governing relation (\ref{eqn:governingRelation}), and observe that there is excellent agreement over a considerable range 
of separations.

\begin{figure}
\begin{center}
\includegraphics[scale=0.5]{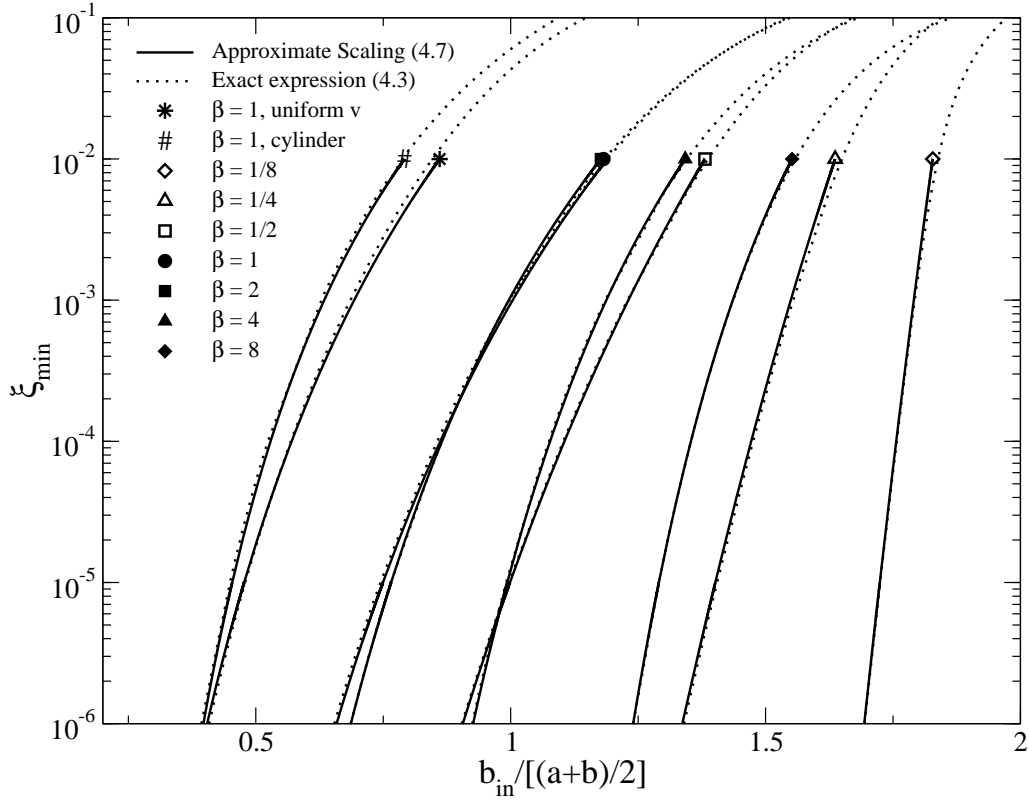}
\end{center}
\caption{Open and filled symbols represent approximate scaling (\ref{eqn:improvedScaling}) compared with full governing 
relation (\ref{eqn:governingRelation}) when a particle is driven by a constant driving force past a spherical obstacle 
for radius-ratios $\beta = 1/8,~1/4,~1/2,~1,~2,~4$ and $8$. (\#) represents a particle being driven past a cylindrical 
obstacle by a constant force, and (*) represents a particle being driven by a uniform velocity field past a spherical 
obstacle, for $\beta = 1$. The multiplicative constant $\lambda_1$ is determined by fitting the data generated with 
(\ref{eqn:governingRelation}) for small values of $\xi$. The fitted values are $1.91,~1.88,~1.84,~1.86,~1.99,~2.33,~2.08$ 
for $\beta=1/8,~1/4,~1/2,~1,~2,~4,~8$, respectively, in the case of the open and filled symbols. For a cylindrical obstacle 
(\#) we obtained $\lambda_1 = 1.58$ and for the case of uniform flow (*) $\lambda_1 = 1.50$.}
\label{fig:scalingEquationCompare}
\end{figure}

\section{Discussion and Summary}\label{sec:hydSurfRghnss}
Using (\ref{eqn:governingRelation}) it is straightforward to extend the concept of
{\it hydrodynamic surface roughness} (henceforth, HR) of spheres introduced by \cite{smart1989}, 
to the motion of a suspended particle past a fixed obstacle. 
In their original work, \citet{smart1989} related the time taken by a sphere
to fall away from a flat surface under the action of gravity to an effective surface roughness. In our case,
we can establish a relationship between the net lateral displacement experienced by a suspended sphere 
moving past an obstacle and the length scale at which surface roughness effects become dominant over
hydrodynamic forces. This relationship is based on the simple but useful
{\it excluded-annulus} model in which the inter-particle potential is
approximated by a hard sphere repulsion with range $(1+\epsilon)a$ \citep{brady1997}.  
In this approximation, the effective roughness determines the minimum separation between particles but
has no effect on the hydrodynamic interaction between them. 
Such hard sphere potentials (in some cases including tangential friction) are widely used to model roughness 
effects on suspension properties, including micro-structure \citep{rampall1997,brady1997,drazer2004,blanc2011}, 
shear-induced dispersion \citep{dacunha1996,drazer2002,ingber2008}, sedimentation \citep{davis1992b,davis2003}, 
rheology \citep{wilson2000,bergenholtz2002}, and transport in periodic systems \citep{frechette2009}.
In our problem, the excluded annulus model implies that, independent of the impact parameter $b_{in}$,
separations smaller than $a \epsilon$ are unattainable due to the hard sphere potential.
On the other hand, the repulsive potential has no effect as the particle separates from the obstacle.
Therefore, if $b_{in}=b_\epsilon$ 
is the incoming impact parameter corresponding to $\xi_{min}=\epsilon$, any trajectory with 
$b_{in}<b_\epsilon$ reaches the same minimum separation $\epsilon$ and
collapses onto the outgoing trajectory corresponding to $b_{in}=b_\epsilon$,
thus inducing a net lateral displacement of magnitude $b_{in}-b_\epsilon$.
In the case of a dilute suspension flowing past a fixed obstacle, this corresponds to the 
presence of a wake of width $2b_{\epsilon}$ behind the obstacle, analogous to that
observed by \cite{khair2006a}. 
The excluded-annulus model implies that the wake behind a
fixed obstacle is related to the HR by (\ref{eqn:governingRelation}),
\begin{equation}
b_{\epsilon} = \left(a+b\right)\left(1+\frac{\epsilon}{2}\right)\exp\left\{H\left(\epsilon\right)\right\}.
\label{eqn:governingRelationRghnss}
\end{equation}
Alternatively, we can view the equation above as the definition of the HR, which serves as an 
effective hard-wall potential resulting from all non-hydrodynamic short-range repulsive interactions. 
The experimental measurement of $b_\epsilon$ would yield such length scale $\epsilon a$ 
for deterministic systems with negligible particle and fluid inertia. A similar surface characterization method has been 
employed in the case of colloidal particles by numerically solving
the particle trajectories \citep{dabros1992, vandeVen1994, wu1996, 
vandeVen1999,whittle2000}. 
We note, however, that due to the presence of a diffusive boundary layer (see \S\ref{subsec:smallXiPDF}), 
(\ref{eqn:governingRelationRghnss}) is valid for sufficiently large P\'{e}clet numbers, such that 
the thickness of the boundary layer is smaller than the repulsive core, i.e., $\xi_{BL}\sim 1/Pe <\epsilon$.

In summary, we have investigated the problem of a spherical, non-Brownian particle negotiating 
a spherical or cylindrical obstacle in the absence of particle and fluid inertia. 
The particle is driven by a uniform ambient velocity field  or a constant force acting on it, and its 
motion is entirely contained in the plane formed by the driving force 
and the radial vector joining the centres of the obstacle and the particle. 
Given this planar nature of the motion and the symmetry of the problem, 
the particle velocity renders itself for decomposition into a radial component (along the centre-to-centre line) 
and a tangential component (perpendicular to the centre-to-centre line), with the
corresponding mobility functions being dependent on the centre-to-centre distance only.
Based on these properties, and extending an approach introduced by \cite{batchelor1972a}
in the context of sheared suspensions, we have derived the steady-state probability distribution 
of particles around the obstacle, assuming a uniform distribution of incoming particles at infinity. 
We showed the distribution to be radially symmetric for both obstacle types and both driving fields, 
analogous to the isotropic pair distribution functions obtained in sheared suspensions, sedimentation 
and microrheology. The asymptotic form of the distribution funtions diverges at contact, suggesting the presence 
of a boundary layer around the obstacle in which Brownian transport is not negligible; existence of such 
a boundary layer has also been reported in the context of sedimenting suspensions and microrheology 
\citep{batchelor1982, nitsche1996, khair2006a}. 
In addition, the asymptotic distribution of particles close to the obstacle is similar for both 
driving fields, and depends only weakly on the dimensionality and geometry of the problem. 
Further, our numerical results indicate that this asymptotic behaviour becomes dominant 
at small separations, highilighting the relevance of other (non-hydrodynamic) interactions.
The other asymptotic limit, that of large separations, yields a divergent excess number 
of particles around the obstacle, suggesting that screening by other particles or container walls 
would eventually become important for the description of the distribution of particles far from the obstacle.

We have also derived an expression for the minimum particle-obstacle separation attained in the course of motion, 
as a function of the incoming impact parameter, using both Eulerian and Lagrangian approaches. We have shown that 
a smaller minimum separation is attained by particles moving in a confined channel, and the separation decreases 
with the extent of the confinement. The asymptotic behaviour in the limit of small impact parameters (particles 
nearly touching the obstacle) shows that the minimum surface-to-surface separation decays exponentially (with a 
negative power of the impact parameter). The exponent governing this asymptotic relationship varies monotonically 
with particle radius, and indicates that for a given obstacle size and sufficiently small incoming impact parameter, 
a smaller particle reaches closer to the obstacle than a larger one. Further, the exponential nature of the 
relationship shows that extremely small surface-to-surface separations can be frequently encountered in the motion 
of particles through an array of obstacles, which, could easily lead to a dominant role of non-hydrodynamic 
interactions in microfluidic systems. Interestingly, the exponential decay of the minimum 
separation as a function of the impact parameter is independent of the dimensionality of the problem and depends 
only weakly on the geometry of the obstacle and aspect ratio.

\begin{acknowledgements}
We would like to thank Profs. A. Acrivos, A. Prosperetti, A. Sangani and J. F. Morris for useful discussions. 
This material is partially based upon work supported by the National Science Foundation under grant no. CBET-0954840.
\end{acknowledgements}

\appendix
\newpage
\section{Mobility fuctions $A$ and $B$}\label{appx:mobFunc}
In this appendix, we describe the radial, $A(r)$, and tangential, $B(r)$, mobility functions used in this work. We also provide the tabulated 
values of the different coefficients used in the expressions provided for the mobility functions at small surface-to-surface separations.
\subsection{Spherical obstacle}\label{appxsubsec:SphObst}
In the case of a spherical obstacle we follow the notation used by \citet{jeffrey1984} and write $A$ and $B$ in 
terms of the scalar functions $x^a_{\alpha\beta}$, $x^b_{\alpha\beta}$, $x^c_{\alpha\beta}$, $y^a_{\alpha\beta}$, $y^b_{\alpha\beta}$ and $y^c_{\alpha\beta}$.
The subscript $\alpha\beta$ denotes a function relating the motion of sphere $\alpha$ to the force or torque acting on sphere $\beta$. 
Consequently the values of $\alpha$ and $\beta$ can be $1$ or $2$. In our case, we use $1$ for the moving particle and $2$ for the obstacle.
The expressions for $A$ and $B$ are given below for both driving fields.
The near-field and far-field approximations presented in \ref{subsec:LubrProblem} and \ref{subsec:mobFuncLargeXi}, respectively,
 were obtained from the equations below using the expressions tabulated by \citet{kim1991}.
 
\subsubsection{Constant force acting on the moving sphere}
\begin{equation}
A = \frac{1}{\mu}\left(\frac{-x^a_{12} x^a_{21} + x^a_{11} x^a_{22}}{x^a_{22}}\right)
\label{eqn27}
\end{equation}

\begin{equation}
B = -\frac{1}{\mu}\left(-y^a_{11} - y^a_{12} \frac{y^a_{21} y^c_{22} - y^b_{21} y^b_{22}}{-y^a_{22} y^c_{22} + (y^b_{22})^2}+ \frac{y^a_{21} y^b_{21}}{y^b_{22}} + \frac{y^a_{22} y^b_{21}}{y^b_{22}}\frac{y^a_{21} y^c_{22} - y^b_{21} y^b_{22}}{-y^a_{22} y^c_{22} + (y^b_{22})^2}\right)
\label{eqn28}
\end{equation}
\vspace{2mm}
In table 1 below, we provide the numerical values of the coefficients $b_i$ and $\gamma_i$ used in \S\ref{subsec:smallXi} for the case of a constant 
force driving the particle:
\begin{center}
\begin{tabular}{cccccccc}
\hline
$\beta$&$b_1$&$b_2$&$b_3$&$b_4$&$\gamma_0$&$\gamma_1$&$\gamma_2$\\
\hline
1/8&1458&31574.8&3115.50&31260.7&0.05991&0.00032&0.09934\\
1/4&250&2900&504.485&2917.17&0.09334&0.00200&0.17093\\
1/2&54&343.805&102.275&348.493&0.11920&0.01013&0.28335\\
1&16&56.2240&28.5301&54.6052&0.11890&0.03778&0.48470\\
2&6.75&12.2487&10.7917&9.67405&0.09389&0.10199&1.01354\\
4&3.90625&2.55981&4.69405&0.45347&0.06189&0.21761&10.1337\\
8&2.84766&-0.7352&1.64041&-1.6223&0.21398&1.4394&0.42824\\
\hline
\end{tabular}
\\\textsc{Table 1.}\hspace{2mm}Coefficients $b_i$ and $\gamma_i$ from (\ref{eqn:newApproxB}) and (\ref{eqn:improvedScaling}). Constant force. Spherical obstacle.
\end{center}

\subsubsection{A freely suspended sphere in a uniform ambient velocity field}
\begin{equation}
A = \frac{1}{6\pi\mu a}\left(1-\frac{x^a_{12}}{x^a_{22}}\right)
\label{eqn29}
\end{equation}

\begin{equation}
B = \frac{1}{6\pi\mu a}\left(1 - \frac{y^b_{21}}{y^b_{22}} + \frac{y^a_{22} y^b_{21} - y^a_{12} y^b_{22}}{y^b_{22}}\frac{y^c_{22}}{y^a_{22} y^c_{22} - (y^b_{22})^2}\right)
\label{eqn30}
\end{equation}
\vspace{2mm}
In table 2, we enlist the values of the coefficients $b_i$ and $\gamma_i$ used in \S\ref{subsec:smallXi} when a freely suspended particle is driven by 
a uniform velocity field:
\begin{center}
\begin{tabular}{cccccccc}
\hline
$\beta$&$b_1$&$b_2$&$b_3$&$b_4$&$\gamma_0$&$\gamma_1$&$\gamma_2$\\
1/8&1452.41&31453.9&3115.50&31260.7&0.06001&0.00032&0.09934\\
1/4&245.43&2847&504.485&2917.17&0.09485&0.00200&0.17093\\
1/2&50.43&321.072&102.275&348.493&0.1285&0.01013&0.28335\\
1&13.18&46.299&28.5301&54.6052&0.1518&0.03778&0.48470\\
2&4.363&7.918&10.7917&9.67405&0.1664&0.10199&1.01354\\
4&1.734&1.1363&4.69405&0.45347&0.1769&0.21761&10.1337\\
8&0.774&-0.1998&1.64041&-1.6223&1.0922&1.4394&0.42824\\
\hline
\end{tabular}
\\\textsc{Table 2.}\hspace{2mm}Coefficients $b_i$ and $\gamma_i$ from (\ref{eqn:newApproxB}) and (\ref{eqn:improvedScaling}). Uniform velocity field. Spherical obstacle.
\end{center}
In table 3, we tabulate the values of the coefficients $k_0$ and $k_1$ defined in \S\ref{subsec:LubrProblem}, in the 
expressions for the mobility functions near contact for the case of a uniform ambient velocity past a spherical obstacle.
\begin{center}
\begin{tabular}{cccccccccccc}
\hline
$\beta$&1/10&1/8&1/5&1/4&1/2&1&2&4&5&8&10\\
$k_0$&0.9965&0.9937&0.9796&0.9661&0.8663&0.6452&0.3648&0.1553&0.1121&0.0532&0.0364\\
$k_1$&0.9978&0.9962&0.9886&0.9817&0.9339&0.8235&0.6464&0.4439&0.3836&0.2718&0.2272\\
\hline
\end{tabular}
\\\textsc{Table 3.}\hspace{2mm}Coefficients $k_0$ and $k_1$ from \S\ref{subsec:LubrProblem}. Uniform velocity field. Spherical obstacle.
\end{center}

\subsection{Cylindrical obstacle}

Unlike the hydrodynamic interactions between two spheres, those between a cylinder and a sphere of comparable radius (external to the cylinder) are 
scantily documented in the literature. \cite{nitsche1996} considers the motion of a sphere near a cylindrical fibre, for the case of a constant force 
acting on the sphere suspended in a quiescent fluid. First, representing the sphere by a point particle and the cylinder by a line of singularities, 
Nitsche obtains the far-field expressions for $A$ and $B$ in (\ref{eqn:nitscheFFA}) and (\ref{eqn:nitscheFFB}).

For small separations between the sphere and the obstacle, the framework established by \cite{cox1974} and generalized by \cite{claeys1989} yields the 
near-field expressions, as documented for the case of a sphere and a cylinder of the same radius (i.e., $\beta=1$) by \cite{nitsche1996}. From these 
expressions, the coefficients $b_i$ and $\gamma_i$ used in \S\ref{subsec:smallXi} are given as follows: $b_1=4.2426,~b_2=0,~b_3=4.6669,~b_4=0$ and 
$\gamma_0=0,~\gamma_1=0.2143$ and $\gamma_2$ is immaterial.

Further, \citet{nitsche1996} provides an approximate functional form for the entire range of separations (for $\beta = 1$) 
using a hyperbolic tangent as a weighting function between the far-field and near-field regions,
\begin{align}\label{eqn:nitscheInterpolatedA}
A(r) &= \frac{1}{6\upi  \mu  a}\left\{\frac{1}{2}\left[1+\text{tanh}\left(2\ln\left(\frac{r/a-2}{1.57}\right)\right)\right]\left[1-\frac{a}{r}\frac{117\upi }{128}\frac{1}{\ln\left(\frac{2r}{a}+1.7\right)}\right]\right.\\ \notag
&~\left.+ \frac{1}{2}6\upi  \left[1-\text{tanh}\left(2\ln\left(\frac{r/a-2}{1.57}\right)\right)\right]\left[\frac{12\upi }{3\sqrt{2}}\frac{1}{(r/a-2)}-\frac{151\upi }{60\sqrt{2}}\ln\left(\frac{3}{4}\left(r/a-2\right)\right)+19.4\right]^{-1}\right.\\ \notag
&~ \left.-0.09\left[\text{cosh}\left(2\ln\left(\frac{r/a-2}{1.57}\right)\right)\right]^{-1}\right\},
\end{align}

\begin{align}\label{eqn:nitscheInterpolatedB}
B(r) &= \frac{1}{6\upi  \mu  a}\left\{\frac{1}{2}\left[1+\text{tanh}\left(2\ln\left(\frac{r/a-2}{1.034}\right)\right)\right]\left[1-\frac{a}{r}\frac{3\upi }{8}\frac{1}{\ln\left(\frac{2r}{a}+2.7\right)}\right]\right.\\ \notag
&~\left.+\frac{1}{2}6\upi  \left[1-\text{tanh}\left(2\ln\left(\frac{r/a-2}{1.034}\right)\right)\right]\left[-\upi\sqrt{2}\ln\left(\frac{3}{4}\left(r/a-2\right)\right)+19.5\right]^{-1}\right.\\ \notag
&~ \left.-0.06\left[\text{cosh}\left(2\ln\left(\frac{r/a-2}{1.034}\right)\right)\right]^{-1}\right\}.
\end{align}
\vspace{2mm}

Finally, we compute the mobility of a sphere in a particle-obstacle system confined between two parallel walls that are perpendicular 
to the axis of the cylinder. To this end, we use an interpolation technique similar to the above expressions, i.e., we interpolate 
between the mobility of a sphere along the mid-plane between two parallel walls \citep{happel1965} and the mobility of a sphere 
in the vicinity of a cylinder:
\begin{align*}
\left(\begin{smallmatrix}\text{effective mobility}\\ \text{in confinement} \end{smallmatrix}\right) &= \frac{1}{2}\left[1+\text{tanh}\left(2\ln\left(\frac{r/a-2}{2\ell_0/a}\right)\right)\right]\left[M_{wall}(\ell_0)\right] \\
 &+ \frac{1}{2}\left[1-\text{tanh}\left(2\ln\left(\frac{r/a-2}{2\ell_0/a}\right)\right)\right]\left[M_{cyl}(r)\right],
\end{align*}
where, \[M_{wall}(r) = \frac{1}{6\pi\mu a}\left[1-1.004\left(\frac{a}{\ell_0}\right)+0.418\left(\frac{a}{\ell_0}\right)^3+0.21\left(\frac{a}{\ell_0}\right)^4-0.169\left(\frac{a}{\ell_0}\right)^5\right],\] and $M_{cyl}(r)\equiv A(r) \text{ or } B(r)$ from (\ref{eqn:nitscheInterpolatedA}) or (\ref{eqn:nitscheInterpolatedB}). Note that $2\ell_0$ -- involved in the hyperbolic tangent functions -- is the separation between the parallel walls. This represents the length-scale at which the interpolation switches from the mobility of a sphere between parallel walls to that in the vicinity of a cylinder.
\vspace{5mm}

\section{Derivation of the equation for the minimum separation}\label{appx:simplificationOfGoverningRelation}

Here we simplify the following equation:
\begin{equation}
\frac{b_{in}^{n-1}}{n-1} =
\displaystyle\int_{a+b}^{r_{min}}\frac{B(r)}{A(r)}r^{n-2}\exp\left\{\displaystyle\int_r^\infty\frac{(n-1)(A-B)}{\tilde{r}
A}d\tilde{r}\right\}dr
\label{eqnB1}
\end{equation}
First, we multiply both sides by $(n-1)$ and then write the {\it RHS} of the equation as:
\begin{eqnarray}
b_{in}^{n-1}= \displaystyle\int_{a+b}^{r_{min}} 
\left[-r^{n-1} \right] \left[ \frac{(n-1)(A-B)}{r A} \exp\left\{\displaystyle\int_r^\infty\frac{(n-1)(A-B)}{\tilde{r}A}d\tilde{r}\right\}\right] dr
\\ \nonumber + \displaystyle\int_{a+b}^{r_{min}} (n-1) r^{n-2} \exp\left\{\displaystyle\int_r^\infty\frac{(n-1)(A-B)}{\tilde{r}A}d\tilde{r}\right\}dr.
\end{eqnarray}
Then, integrating the first term by parts we obtain,
\begin{equation}
b_{in}^{n-1} = \left(r_{\min }\right)^{n-1}\exp\left\{\displaystyle\int_{r_{\min }}^{\infty}\frac{(n-1)(A-B)}{\tilde{r}A}d\tilde{r}\right\}-(a+b)^{n-1}\exp\left\{\displaystyle\int_{a+b}^{\infty}\frac{(n-1)(A-B)}{\tilde{r}A} d\tilde{r}\right\}
\label{eqn:appB2}
\end{equation}

The last term above can be written formally as the following limit,
\begin{align}\label{eqn:appB3}
\lim_{\xi \to 0} \left[\left(a+b\right)\left(1+\frac{\xi}{2}\right)\right]^{n-1} \exp\left\{(n-1) H\left(\xi \right)\right\} =0
\end{align}
Where $H(\xi)$ was defined in (\ref{eqn:definitionOfH}) of \S\ref{subsec:smallXiPDF} and the limiting behavior can be obtained using the near-field lubrication expressions for $A$ and $B$. 
Therefore, we get
\begin{equation}
b_{in}^{n-1}=r_{min}^{n-1}\exp\left\{\displaystyle\int_{r_{min}}^\infty\frac{(n-1)(A-B)}{\tilde{r}
A}d\tilde{r}\right\}
\label{eqnB2}
\end{equation}
Taking $(n-1)^{th}$ root on both sides, one obtains the required relation (\ref{eqn:governingRelation}) in terms of $r$. Simplification of this equation in terms of dimensionless separation is obtained by substituting $\xi = 2\left(r-a-b\right)/(a+b)$.

\bibliographystyle{jfm}

\end{document}